\documentstyle[12pt]{article}
\begin{document}
\begin{titlepage}
\begin{flushright}
IC/2001/9\\
hep-th/0102194
\end{flushright}
\vspace{10 mm}

\begin{center}
{\Large Variable-Speed-of-Light Cosmology from Brane World Scenario}

\vspace{5mm}

\end{center}
\vspace{5 mm}

\begin{center}
{\large Donam Youm\footnote{E-mail: youmd@ictp.trieste.it}}

\vspace{3mm}

ICTP, Strada Costiera 11, 34014 Trieste, Italy

\end{center}

\vspace{1cm}

\begin{center}
{\large Abstract}
\end{center}

\noindent

We argue that the four-dimensional universe on the TeV brane of the 
Randall-Sundrum scenario takes the bimetric structure of Clayton and Moffat, 
with gravitons traveling faster than photons instead, while the radion varies 
with time.  We show that such brane world bimetric model can thereby solve 
the flatness and the cosmological constant problems, provided the speed of a 
graviton decreases to the present day value rapidly enough.  The resolution 
of other cosmological problems such as the horizon problem and the monopole 
problem requires supplementation by inflation, which may be achieved by the 
radion field provided the radion potential satisfies the slow-roll 
approximation.

\vspace{1cm}
\begin{flushleft}
February, 2001
\end{flushleft}
\end{titlepage}
\newpage

\section{Introduction}\label{one}

Variable-Speed-of-Light (VSL) cosmological models were proposed \cite{mof1,am} 
as an alternative to inflation \cite{gut,lin,als} for solving the cosmological 
problems of the Standard Big Bang (SBB) model.  VSL models assume that the 
speed of light initially took a larger value and then decreased to the present 
value at an early time.  Although the basic idea may be controversial and 
the theoretical foundation is not yet well-developed, VSL models are  
appealing in that not only the cosmological problems solved by the 
inflationary models but also the cosmological constant problem can be solved 
\cite{mof1,am,bar1,bar2,bar3,mof2,bar4}.  The claim in Ref. \cite{wfc} of the 
experimental evidence for a time-varying fine structure constant $\alpha=
e^2/(4\pi\hbar c)$ suggests that the speed of light may indeed vary with 
time
\footnote{Other possibility of time-varying electric charge $e$ was considered 
in Ref. \cite{bek}.}.  
More recent observational evidence for a time-varying fine structure constant 
can be found, for example, in Refs. \cite{kst,bir,bcw,amrv,mur,web,ave2}.    
Furthermore, the recent works 
\cite{kal1,kir,kal2,chu,ale,ish,ckr,csa} on the Lorentz violation in the brane 
world scenario hint at the possibility of naturally realizing VSL models 
within brane world scenario.  

In the original models by Moffat \cite{mof1} and by Albrecht and Magueijo 
\cite{am}, the speed of light (in the action), a fundamental constant of 
nature, is just assumed to vary with time during an early period of cosmic 
evolution and thereby the Lorentz symmetry becomes explicitly broken.  
It is therefore assumed in such models that there exists a preferred 
frame in which the laws of physics take standard forms (of the Lorentz 
invariant theories) with the constant speed of light $c$ replaced by a 
field $c=c(x^{\mu})$, the so-called principle of minimal coupling.  Although 
many cosmological problems can be solved through such approach, such radical 
modification of standard physics may be questionable.  In the previous paper 
\cite{youm}, we applied the approach of Refs. \cite{mof1,am,bar1,bar3} to the 
cosmological models of the Randall-Sundrum (RS) scenario \cite{rs1,rs2,rs3} 
for the purpose of studying its consequences.  

Clayton and Moffat \cite{cm1,cm2} proposed an ingenious dynamical mechanism 
by which the speed of light can vary with time in a diffeomorphism invariant 
manner and without explicitly breaking the Lorentz symmetry.  (See also Ref. 
\cite{dru} for an independent development.)  Their models therefore avoid 
the need to introduce a global preferred frame into spacetime.  They 
introduce two metrics into the spacetime manifold, one being associated 
with gravitons and the other providing the geometry on which matter fields, 
including photons, propagate.  These two metrics are nonconformally related 
by a scalar field (called a biscalar) or a vector field (called a bivector).  
The causal structures determined by the two metrics are therefore different, 
thereby photons propagate at different speed from that of gravitons.  

In this paper, we argue that the bimetric mechanism by Clayton and Moffat 
can be naturally realized within the brane world scenarios.  If we define 
the radion as the distance between the two branes, rather than in terms 
of the extra spatial component of the bulk metric, then the radion 
nonconformally relates the induced metric on the TeV brane, to which matter 
fields on the TeV brane are minimally coupled, to the gravity metric of the 
TeV brane in the same manner in which a biscalar does, while the radion 
varies with time.  So, the radion can be regarded as a biscalar.  
Since the biscalar term in the induced metric comes with the opposite 
sign from that of Ref. \cite{cm2}, gravitons rather appear to propagate 
{\it faster} than photons on the TeV brane before the radion is stabilized, 
rather than photons traveling faster than gravitons, as was the case in Ref. 
\cite{cm2}.   This difference allows the bimetric model resulting from the 
brane world scenario to solve the flatness and the cosmological constant 
problems (which the original bimetric models were not able to solve by 
itself), provided that the radion potential has an appropriate form giving 
rise to rapid enough decrease of the speed of gravitons to the present value.  
When the speed of a graviton does not decrease rapidly enough, the quasi 
flatness and the quasi cosmological constant problems are expected to be 
solved instead.  On the other hand, since the speed of a photon remains 
constant with the natural choice of the time coordinate for the TeV brane 
observer (or decreases with the choice of the comoving time coordinate of 
the gravity metric), our bimetric model cannot by itself solve other 
cosmological problems, such as the horizon problem and the monopole problem, 
which the original VSL models can solve.  To solve such problems, our brane 
world bimetric model has to be supplemented by inflation.  Provided the radion 
potential has a region satisfying the slow-roll approximation, the radion may 
also be used as an inflaton.  Although our bimetric model may turn out to be 
insufficient for solving the cosmological problems, it can at least provide 
with a brane world scenario explanation for a time-varying fine structure 
constant observed in our universe.  

The paper is organized as follows.  In section \ref{two}, we derive the 
effective Friedmann equations for cosmological model of the RS1 model and 
show that gravitons and photons travel at different speeds while the radion 
varies with time.  In sections \ref{three} and \ref{four}, we argue that our 
bimetric model can solve the flatness and the cosmological constant problems, 
provided the speed of gravitons decreases to the present value rapidly enough. 
In section \ref{five}, we comment on other cosmological problems.

\section{Effective Friedmann Equations with Time-Varying Radion}\label{two}

In this section, we obtain the effective Friedmann equations on the TeV brane 
while the radion varies with time.  The details on derivation can be found 
in Ref. \cite{cgrt} (see also Ref. \cite{cf2}), which we follow closely.  
However, we rederive the equations since we choose to define the radion 
differently from Ref. \cite{cgrt}.  

We define the extra spatial coordinate $y$ such that one of the branes, 
which we choose to be the Planck brane, is at rest at $y=0$.  In such 
coordinate system, the TeV brane moves along the $y$-direction before  
the radion gets stabilized.  We choose to encode the distance between 
the two branes entirely with the location $y=R(t)$ of the TeV brane. 
With this convention, the radion is defined as the relative distance 
$R(t)$ between the two branes instead of the metric component along 
the extra dimension.  Namely, the time evolution of the size of the extra 
space is described not by that of the extra spatial component of the bulk 
metric, but by that of the location $y=R(t)$ of the TeV brane.  
This coordinate choice was also previously considered in Ref. \cite{bdl} 
for the purpose of studying the radion dynamics in brane cosmology.

The general Ansatz for the bulk metric describing the expanding brane 
universe in our convention is given by
\begin{equation}
\hat{g}_{MN}dx^Mdx^N=-n^2(t,y)c^2dt^2+a^2(t,y)\gamma_{ij}dx^idx^j+dy^2,
\label{bulkmet}
\end{equation}
where $\gamma_{ij}$ is the metric for the maximally symmetric 
three-dimensional space given in the Cartesian and the spherical coordinates 
by
\begin{equation}
\gamma_{ij}dx^idx^j=\left(1+{k\over 4}\delta_{mn}x^mx^n\right)^{-2}\delta_{ij}
dx^idx^j={{dr^2}\over{1-kr^2}}+r^2(d\theta^2+\sin^2\theta d\phi^2),
\label{mxsymmet}
\end{equation}
with $k=-1,0,1$ for the three-dimensional space with the negative, zero and 
positive spatial curvature, respectively.   
The tensions of the Planck and the TeV branes are respectively denoted as 
$\sigma_1$ and $\sigma_2$.  The Planck and the TeV branes contain matter 
fields with the Lagrangian densities ${\cal L}_1$ and ${\cal L}_2$, 
respectively.  We assume that the radion potential $V_r(R)$ has been generated 
through some mechanism such as the Goldberger-Wise mechanism \cite{gw1,gw2}. 

When there is no matter on the branes, the equations of motion are solved 
by the static brane solution \cite{rs1,rs2} with the metric components
\begin{equation}
n(y)=a(y)=e^{-m_0|y|},\ \ \ \ \ \ \ \ \  \gamma_{ij}=\delta_{ij}.
\label{statsol}
\end{equation}
The brane tensions take the fine-tuned values given by
\begin{equation}
\sigma_1={{3c^4m_0}\over{4\pi G_5}}=-\sigma_2,\ \ \ \ \ \ \ \ \ \ \ 
\Lambda=-{{3c^4m^2_0}\over{4\pi G_5}}.
\label{brntens}
\end{equation}

When matter fields are included on the branes, the brane universe undergoes 
cosmological evolution.  Since the solution for the expanding brane universe 
should reduce to the static solution (\ref{statsol}) in the limit of vanishing 
mass densities $\varrho_{1,2}$ and pressures $\wp_{1,2}$ of the matter fields,
it is reasonable to parametrize the solution in terms of the linear expansion 
around the static solution in the following way:
\begin{eqnarray}
n(t,y)&=&\Omega(y)\left(1+\delta n(t,y)\right),
\cr
a(t,y)&=&a(t)\Omega(y)\left(1+\delta a(t,y)\right),
\label{expmetrc}
\end{eqnarray}
with the perturbations $\delta n$ and $\delta a$ assumed to be of the order 
of $\varrho_{1,2}$.  Here, $\Omega(y)\equiv e^{-m_0|y|}$.  To obtain the 
effective Friedmann equations describing the expanding brane universe as 
observed on the brane, we either take the averages of the Einstein's 
equations as $\int_0^{R(t)}dy\,\Omega^4{\cal G}^M_N={{8\pi G_5}\over{c^4}}
\int^{R(t)}_0dy\,\Omega^4{\cal T}^M_N$ or compute the four-dimensional 
effective action, with the linear perturbations (\ref{expmetrc}) substituted 
\cite{cgrt}.  In doing so, we keep only terms leading order in 
$\varrho_{1,2}$.  

Following the section 4.1 of Ref. \cite{cgrt}, we drop the $\delta n$ and 
$\delta a$ perturbations when calculating the effective action, as they 
contribute at ${\cal O}(\varrho^2_{1,2})$.  Namely, we consider the following 
form for the bulk metric:
\begin{equation}
n(y)=e^{-m_0|y|},\ \ \ \ \ \ \ \ 
a(t,y)=a(t)e^{-m_0|y|}.
\label{ldorbm}
\end{equation}
The induced metrics on the Planck and the TeV branes are then respectively
\begin{equation}
g_{1\,\mu\nu}dx^{\mu}dx^{\nu}=-c^2dt^2+a^2(t)\gamma_{ij}dx^idx^j,
\label{plindmet}
\end{equation}
\begin{equation}
g_{2\,\mu\nu}dx^{\mu}dx^{\nu}=-\left[\Omega^2_Rc^2-\dot{R}^2\right]dt^2
+a^2(t)\Omega^2_R\gamma_{ij}dx^idx^j,
\label{tevindmet}
\end{equation}
where $\Omega_R\equiv\Omega(R(t))=e^{-m_0R(t)}$ and the overdot stands for 
derivative w.r.t. $t$.  The four-dimensional effective action therefore takes 
the form
\begin{eqnarray}
S_{eff}&=&{{3c^3}\over{8\pi G_5m_0}}\int dx^0\,\sqrt{\gamma}a^3(1-\Omega^2_R)
\left[{\dot{a}^2\over a^2}+{\ddot{a}\over a}+{{kc^2}\over a^2}\right]
+c\int dx^0\,\sqrt{\gamma}a^3V_r(R)
\cr
& &+c\int dx^0\,\sqrt{\gamma}a^3{\cal L}_1
+c\int dx^0\,\sqrt{\gamma}a^3\sqrt{\Omega^2_R-\dot{R}^2/c^2}
\Omega^3_R{\cal L}_2,
\label{effact}
\end{eqnarray}
where $x^0=ct$ and $\gamma\equiv{\rm det}\gamma_{ij}$.  Noting that the Ricci 
scalar for the four-dimensional Robertson-Walker metric $g_{\mu\nu}dx^{\mu}
dx^{\nu}=-c^2dt^2+a(t)^2\gamma_{ij}dx^idx^j$ is given by 
\begin{equation}
{\cal R}={6\over c^2}\left[{\ddot{a}\over a}+\left({\dot{a}\over a}\right)^2
+{{kc^2}\over a^2}\right],
\label{4dricscal}
\end{equation}
we can put the above effective action in a conventional looking form:
\begin{equation}
S_{eff}=\int dx^0\sqrt{-g}\left[{c^4\over{16\pi G_{eff}}}{\cal R}+V_r(R)+
{\cal L}_1+\sqrt{\Omega^2_R-\dot{R}^2/c^2}\Omega^3_R{\cal L}_2\right],
\label{conveffact}
\end{equation}
where $g=\det g_{\mu\nu}$ and $G_{eff}\equiv m_0G_5/(1-\Omega^2_R)$.  
Therefore, we see that $a(t)$ will satisfy the four-dimensional Friedmann 
equations with the contribution to the mass density and the pressure coming 
from ${\cal L}_{1,2}$.  

To obtain the effective Friedmann equations satisfied by $a(t)$ from the 
effective action (\ref{conveffact}), we have to define the mass densities 
$\varrho_{1,2}$ and the pressures $\wp_{1,2}$ of the matter fields (described 
by the Lagrangian densities ${\cal L}_{1,2}$) on the Planck and the TeV 
branes.  The energy-momentum tensors for the matter fields on the branes are 
defined as 
\begin{equation}
{\cal T}^{\mu\nu}_{1,2}=-{2\over\sqrt{-g_{1,2}}}{{\delta(\sqrt{-g_{1,2}}
{\cal L}_{1,2})}\over{\delta g_{1,2\,\mu\nu}}},
\label{setens}
\end{equation}
where $g_{1,2}$ are determinants of the induced metrics (\ref{plindmet},
\ref{tevindmet}) on the Planck and the TeV branes.  Note, we defined the 
energy-momentum tensors in terms of the induced metrics $g_{1,2\,\mu\nu}$, 
since the induced metrics are the physical metrics for the matter fields on 
the branes.  Modeling the matter fields as perfect-fluid, we can put these 
energy-momentum tensors into the following standard forms:
\begin{equation}
{\cal T}^{\mu\nu}_{1,2}=\left(\varrho_{1,2}+{\wp_{1,2}\over c^2}\right)
U^{\mu}_{1,2}U^{\nu}_{1,2}+\wp_{1,2}g^{\mu\nu}_{1,2}.
\label{pfsetens}
\end{equation} 
Since the four-velocities $U^{\mu}_{1,2}$ of the fluid on the Planck and 
the TeV branes are normalized as $g_{1,2\,\mu\nu}U^{\mu}_{1,2}U^{\nu}_{1,2}
=-c^2$, the nonzero components of $U^{\mu}_{1,2}$ in the comoving coordinates 
are given by
\begin{equation}
U^t_1=1,\ \ \ \ \ \ \ \ \ 
U^t_2={1\over\sqrt{\Omega^2_R-\dot{R}^2/c^2}}.
\label{nzcpus}
\end{equation}
So, the nonzero components of the energy-momentum tensors for the matter 
fields on the Planck and the TeV branes are respectively
\begin{equation}
{\cal T}^{tt}_1=\varrho_1,\ \ \ \ \ \ \ \ \ \ \ \ 
{\cal T}^{ij}_1={\wp_1\over{a^2}}\gamma^{ij},
\label{nzcpemtens1}
\end{equation}
\begin{equation}
{\cal T}^{tt}_2={\varrho_2\over{\Omega^2_R-\dot{R}^2/c^2}},\ \ \ \ \ \ \
{\cal T}^{ij}_2={\wp_2\over{a^2\Omega^2_R}}\gamma^{ij}.
\label{nzcpemtens2}
\end{equation}
For the purpose of putting the Friedmann equations into simple and 
suggestive forms, we reparametrize the mass density and the pressure of 
the matter fields on the Planck brane in the following way:
\begin{equation}
\varrho_1={{\tilde{\varrho}_1\Omega_R}\over{\sqrt{\Omega^2_R-\dot{R}^2/c^2}}},
\ \ \ \ \ \ \ \ \ \ 
\wp_1={{\tilde{\wp}_1\sqrt{\Omega^2_R-\dot{R}^2/c^2}}\over\Omega_R}.
\label{reparam}
\end{equation}
After the radion is stabilized, i.e., $\dot{R}=0$, we have $\tilde{\varrho}_1
=\varrho_1$ and $\tilde{\wp}_1=\wp_1$.  
Making use of these facts, we obtain the following effective Friedmann 
equation:
\begin{equation}
\left({\dot{a}\over a}\right)^2+{{kc^2}\over a^2}=
{{8\pi m_0G_5}\over{3(1-\Omega^2_R)}}\left[{{\tilde{\varrho}_1\Omega_R
+\varrho_2\Omega^5_R}\over\sqrt{\Omega^2_R-\dot{R}^2/c^2}}+V_r/c^2\right].
\label{effecredeq}
\end{equation}

Note, the physical metric for the matter fields on the TeV brane, on which 
we are assumed to live, is given by the induced metric (\ref{tevindmet}).  
So, the cosmic scale factor on the TeV brane is actually given by $a_R=
a\Omega_R$.  In terms of the physical cosmic scale factor $a_R$, the 
effective Friedmann equation (\ref{effecredeq}) takes the form:
\begin{equation}
\left({\dot{a}_R\over a_R}\right)^2+2m_0\dot{R}{\dot{a}_R\over a_R}+m^2_0
\dot{R}^2+{{kc^2\Omega^2_R}\over a^2_R}={{8\pi m_0G_5}\over{3(1-\Omega^2_R)}}
\left[{{\tilde{\varrho}_1\Omega_R+\varrho_2\Omega^5_R}\over\sqrt{\Omega^2_R-
\dot{R}^2/c^2}}+V_r/c^2\right].
\label{effecredeq1}
\end{equation}
By defining the cosmic time $\tau$ of the TeV brane in the following way, 
\begin{equation}
d\tau^2\equiv \left[\Omega_R^2-\dot{R}^2/c^2\right]dt^2,
\label{tevcostm}
\end{equation}
we can bring the induced metric (\ref{tevindmet}) on the TeV brane into the 
following standard comoving frame form for the Robertson-Walker metric:
\begin{equation}
g_{2\,\mu\nu}dx^{\mu}dx^{\nu}=-c^2d\tau^2+a^2_R(\tau)\gamma_{ij}dx^idx^j.
\label{tevrwmet}
\end{equation}
In terms of the cosmic time $\tau$, the effective Friedmann equation 
(\ref{effecredeq1}) takes the form:
\begin{eqnarray}
\left({\dot{a}_R\over a_R}\right)^2+2m_0\dot{R}{\dot{a}_R\over a_R}+m^2_0
\dot{R}^2+{{k(c^2+\dot{R}^2)}\over a^2_R}=
\ \ \ \ \ \ \ \ \ \ \ \ \ \ \ \ \ \ \ \ \ \ \ \ \ \ \ \ \ \ \ \ 
\cr
{{8\pi m_0G_5(1+\dot{R}^2/c^2)^{3/2}}\over{3\Omega^2_R(1-\Omega^2_R)}}
\left[\tilde{\varrho}_1+\varrho_2\Omega^4_R+{V_r\over{c^2\sqrt{1+
\dot{R}^2/c^2}}}\right],
\label{effecredeq2}
\end{eqnarray}
where the overdot now stands for derivative w.r.t. $\tau$.  

We have to keep in mind that the parameter $c$ appearing in the conventional 
Friedmann equations (through the terms $kc^2/a^2$ and $\wp/c^2$) corresponds 
to the speed of gravitons for the obvious reason that the metric, in which 
the same $c$ appears, describes the geometry on which {\it gravitons} 
propagate.  So, we see from Eq. (\ref{effecredeq2}) that the effective speed 
of gravitons $c_4$, as well as the effective four-dimensional Newton's 
constant $G_4$, is observed to change with time on the TeV brane in the 
following way:
\begin{equation}
c_4=\sqrt{c^2+\dot{R}^2},\ \ \ \ \ \ \ \ \ \ \ \ \ 
G_4={{m_0G_5(1+\dot{R}^2/c^2)^{3/2}}\over{\Omega^2_R(1-\Omega^2_R)}}.
\label{effeccg}
\end{equation}
The speed of a graviton is therefore observed on the TeV brane to take a 
larger value than the present day value while the radion varies with time.  
So, there will be a time lag between transmission of gravitational waves and 
that of photons along the null surfaces of their respective physical metrics, 
while the radion varies with time.  After the radion is stabilized, the speed 
of a graviton decreases to the present value, coinciding with the speed of 
a photon.  This is expected to be a generic feature of any brane world 
scenarios involving two branes.

In the Friedmann equation (\ref{effecredeq2}), the radion $R$ is mixed with 
the massless graviton through the interaction term $\sim\dot{R}\dot{a}_R$.  
To separate the fields, we perform the conformal transformation on the metric:
\begin{equation}
a_R(\tau)=e^{m_0(R_0-R)}\bar{a}_R,\ \ \ \ \ \ \ \ \ \ \ 
d\tau=e^{m_0(R_0-R)}d\bar{\tau},
\label{cnftrmet}
\end{equation} 
where $R_0$ denotes the present day value of $R$.  The difference between 
the new frame and the original frame is important only for large departure of 
$R$ from $R_0$.  In this new frame, the effective Friedmann equation 
(\ref{effecredeq2}) on the TeV brane takes the form:
\begin{equation}
\left({\dot{\bar{a}}_R\over\bar{a}_R}\right)^2+{{k(c^2+R^2_\tau)}\over
\bar{a}^2_R}={{8\pi m_0e^{2m_0R_0}G_5(1+R^2_\tau/c^2)^{3/2}}\over
{3(1-\Omega^2_R)}}\left[\tilde{\varrho}_1+\varrho_2\Omega^4_R+
{V_r\over{c^2\sqrt{1+R^2_{\tau}/c^2}}}\right],
\label{effecredeq3}
\end{equation}
where the overdot from now on stands for derivative w.r.t. $\bar{\tau}$ and 
the subscript $\tau$ denotes derivative w.r.t. $\tau$.  So, in this new 
frame the effective speed of gravitons $\bar{c}_4$ remains the same but the 
effective Newton's constant $\bar{G}_4$ is rescaled:
\begin{equation}
\bar{c}_4=\sqrt{c^2+R^2_\tau},\ \ \ \ \ \ \ \ \ \ 
\bar{G}_4={{m_0e^{2m_0R_0}G_5(1+R^2_\tau/c^2)^{3/2}}\over{1-\Omega^2_R}},
\label{effredeq3}
\end{equation}
which is expected as properties of conformal transformation.
The other Friedmann equation in this new frame is given by
\begin{equation}
{\ddot{\bar{a}}_R\over\bar{a}_R}=-{{4\pi\bar{G}_4}\over 3}\left[\varrho+
{3\over\bar{c}^2_4}\wp-{{2V_r}\over{c^2\sqrt{1+R^2_{\tau}/c^2}}}\right],
\label{fredeq2}
\end{equation}
where
\begin{equation}
\varrho\equiv\tilde{\varrho}_1+\varrho_2\Omega^4_R,
\ \ \ \ \ \ \ \ \ \ \ \ 
\wp\equiv\tilde{\wp}_1+\wp_2\Omega^4_R.
\label{defrhonwp}
\end{equation}
From now on, for the purpose of simplifying the equations, we shall assume 
that the radion potential terms in the above effective Friedmann equations 
are absorbed into $\varrho$ and $\wp$, i.e., $\varrho\to\varrho+V$ and $\wp
\to\wp-V/\bar{c}^2_4$ ($V\equiv V_r/(c^2\sqrt{1+R^2_{\tau}/c^2})$), and the 
radion $R$ varies with time as specified by the radion potential.

We have therefore shown that the speed of gravitons, as well as the 
Newton's constant, is observed to vary with time on the TeV brane while 
the radion field varies with time.  This fact can be understood as follows.  
Gravitons propagate in the bulk whose geometry is described by the bulk 
metric $\hat{g}_{MN}$ given by Eq. (\ref{bulkmet}).  Therefore, on the TeV 
brane, gravitons are observed to propagate on the geometry described by 
\begin{equation}
g^{grav}_{\mu\nu}dx^{\mu}dx^{\nu}=\left.\hat{g}_{\mu\nu}
\right|_{y=R(t)}dx^{\mu}dx^{\nu}\approx-\Omega^2_Rc^2dt^2+a^2\Omega^2_R
\gamma_{ij}dx^idx^j.  
\label{gravmet}
\end{equation}
On the other hand, the physical metric to which matter fields, including 
photons, on the TeV brane are minimally coupled is given by the induced 
metric (\ref{tevindmet}) on the TeV brane.  These two metrics are 
nonconformally related when $\dot{R}\neq 0$ and coincide when $\dot{R}=0$.  
Thereby, the four-dimensional universe on the TeV brane takes bimetric 
structure, proposed by Clayton and Moffat \cite{cm1,cm2}, with the radion 
$R$ identified as a biscalar.  The four-velocity vector $v^{\nu}$ of a 
photon, which is null w.r.t. the induced metric $g_{2\,\mu\nu}$, i.e., 
$g_{2\,\mu\nu}v^{\mu}v^{\nu}=0$, is timelike
\footnote{Note, for the bimetric models of Refs. \cite{cm1,cm2} the 
four-velocity vector of a photon is {\it spacelike} w.r.t. the gravitational 
metric, because of the difference in the sign in the biscalar term of the 
matter metric.} 
w.r.t. the gravitational metric, i.e., $g^{grav}_{\mu\nu}v^{\mu}v^{\nu}=
-(v^t\partial_t R)^2<0$ when $\partial_tR\neq 0$.  Gravitons therefore 
appear to propagate faster than photons on the TeV brane before the radion 
is stabilized.  As pointed out in the above, the parameter $c$ appearing in 
the conventional Friedmann equations corresponds to the speed of gravitons.  
On the other hand, $c$ appearing in the Robertson-Walker metric 
(\ref{tevrwmet}) corresponds rather to the speed of a photon, because this 
metric describes geometry on which matter fields, including photons, on the 
TeV brane propagates.  This $c$ is the same $c$ that appears in the effective 
Friedmann equations (\ref{effecredeq3},{\ref{fredeq2}).  Since the 
four-velocity vectors $v^{\mu}_g$ and $v^{\mu}_p$ for a graviton and a photon 
are null w.r.t. $g^{grav}_{\mu\nu}$ and $g_{2\,\mu\nu}$, respectively, the 
ratio of the speed of a graviton $c_g$ to the speed of a photon $c_p$ is
\begin{equation}
{c_g\over c_p}={\Omega_R\over\sqrt{\Omega^2_R-(\partial_tR)^2/c^2}}=
\sqrt{1+(\partial_{\tau}R)^2/c^2},
\label{ratio}
\end{equation}
which leads to the same expression for the effective speed of a graviton on 
the TeV brane given in Eqs. (\ref{effeccg},\ref{effredeq3}) upon 
identifying $c_p=c$.  Indeed, the gravitational metric (\ref{gravmet}) 
expressed in terms of the cosmic time $\tau$ for the matter metric 
(\ref{tevrwmet}) takes the following comoving coordinate form with the 
time-varying speed of a graviton $c_g=c_4$ given by Eq. (\ref{effeccg}):
\begin{equation}
g^{grav}_{\mu\nu}dx^{\mu}dx^{\nu}=-c^2_4(\tau)d\tau^2+a^2_R(\tau)\gamma_{ij}
dx^idx^j.
\label{cmvgrvmet}
\end{equation} 

Note that having a constant speed of a photon but a time-varying speed of 
a graviton is a frame-dependent statement.  Since we choose to measure 
speeds with the time coordinate $\tau$ or $\bar{\tau}$, for which the 
constant $c$ corresponds to the speed of a photon, we should regard the 
speed of a graviton as changing with time and taking larger value than the 
speed of a photon before the radion is stabilized.  Had we chosen to use 
the time coordinate (defined by $d\tau^2\equiv\Omega^2_Rdt^2$) which brings 
the gravity metric (\ref{gravmet}) to the comoving frame form to measure 
speeds, the constant $c$ would have corresponded to the speed of a graviton, 
and the speed of a photon should be regarded as changing with time and taking 
smaller value than the speed of graviton before the radion is stabilized.  
Indeed, with a choice of such time coordinate the resulting effective 
Friedmann equations will have constant speed of graviton.

We have mapped the bimetric model resulting from the brane world scenario to  
a model with varying fundamental constants proposed in Refs. \cite{mof1,am}, 
in the sense that the parameter $c$ in the Friedmann equations that is 
assumed to take a large value at an early period in Refs. \cite{mof1,am} is 
actually the speed of {\it a graviton}, not the speed of a photon.  So, our 
bimetric model can resolve the flatness and the cosmological constant 
problems, which are shown to be resolved in Refs. \cite{mof1,am} through 
rapid enough decreasing $c$ in the Friedmann equations.  On the other hand, 
the resolution of some of cosmological problems, such as horizon problem, 
through the VSL models requires a larger value of the speed of {\it a photon} 
at an early period of cosmic evolution.  So, strictly speaking, the bimetric 
VSL models cannot by itself solve all the cosmological problems that were 
originally claimed \cite{mof1,am} to be solved by the VSL models with 
time-varying fundamental constants, since resolution of some of the problems 
requires faster speed of a graviton (namely, a larger $c$ in the Friedmann 
equations) and the others faster speed of a photon.  Nevertheless, our 
bimetric model has an advantage over the previous bimetric models, in the 
sense that the only cosmological problem that the VSL models are claimed to 
solve but inflaton cannot is the cosmological constant problem, which 
requires larger value of speed of a graviton at an early time.  When 
supplemented with inflation, our bimetric model therefore has potential of 
solving all the cosmological problems that are originally claimed to be 
solved by VSL models.  

In Ref. \cite{cou} it was pointed out that the VSL cosmology 
\cite{mof1,am,bar1,bar2,bar3,mof2} has limited ability to resolve the 
Planck problem and can make it worse, because a variable speed of light 
affects the Planck scale.  Strictly speaking, $c$ appearing in the Planck 
mass $m_{pl}=\sqrt{\hbar c/G}$, the Planck length $l_{pl}=\sqrt{\hbar G/c^3}$ 
and the Planck time $t_{pl}=\sqrt{\hbar G/c^5}$ are speed of graviton, 
because these quantities have to do with the quantum gravity effect.  In 
our bimetric model, also the speed of graviton varies with time, taking 
larger value than the (constant) speed of photon, while the radion varies 
with time.  However, for our case, the effective four-dimensional Newton's 
constant, given in Eq. (\ref{effredeq3}), also varies with time, taking {\it 
larger} value than the present value.  With the effective four-dimensional 
speed of graviton and Newton's constant given in Eq. (\ref{effredeq3}) 
substituted, the Planck mass $m_{pl}$ takes {\it smaller} value than the 
present value, the Planck length $l_{pl}$ remains constant and the decrease 
in the Planck time $t_{pl}$ becomes less severe than the previous VSL 
cosmological models with varying fundamental constants, while the radion 
varies with time.  So, first of all, our bimetric model does not make the 
hierarchy problem worse, unlike the previous VSL models with varying 
fundamental constants.  However, the Planck density $\sim m_{pl}/l^3_{pl}$ 
rather decreases in our case, so our bimetric model by itself cannot resolve 
the Planck density problem.  

From the effective Friedmann equations (\ref{effecredeq3},\ref{fredeq2}) 
(of course, with the radion potential terms absorbed into $\varrho$ and 
$\wp$), we obtain the following generalized conservation equations:
\begin{equation}
\dot{\varrho}+3\left(\varrho+{\wp\over\bar{c}^2_4}\right){\dot{\bar{a}}_R
\over\bar{a}_R}=-\varrho{\dot{\bar{G}}_4\over\bar{G}_4}+{{3k\bar{c}_4
\dot{\bar{c}}_4}\over{4\pi\bar{G}_4\bar{a}^2_R}}.
\label{genconseq}
\end{equation}
So, with the time-varying radion field $R$, which causes the time-variation 
of the effective speed of gravitons and the effective Newton's constant, 
mass is not conserved on the TeV brane, implying that matter is created in 
the brane universe.   Even if the comoving time coordinate for the 
gravitational metric $g^{grav}_{\mu\nu}$ were chosen, the mass appears to 
be not conserved on the TeV brane due to the time variation of the effective 
Newton's constant.  This result is in contrast with the bimetric models 
of Refs. \cite{cm1,cm2}, for which the mass density of the ordinary matter 
is conserved.  This difference may be attributed to the following reason.  
In Refs. \cite{cm1,cm2}, it is assumed from the outset that the 
energy-momentum tensor for the ordinary matter satisfies the conservation 
law (w.r.t. the matter metric) and then shown that the conservation law is 
consistent with the field equations and the Bianchi identities.  In our case, 
the conservation law for the matter fields on the branes is not compatible 
with the four-dimensional effective theory while the radion field varies with 
time.  This is due to the fact that the four-dimensional effective Newton's 
constant (and the speed of gravitons with a suitable choice of frame) in the 
four-dimensional effective action is time-dependent while the radion 
varies with time.  As was elaborated in Refs. \cite{mof1,am}, the 
energy-momentum tensor conservation law is incompatible with the Bianchi 
identities  ${\cal G}^{\mu\nu}_{\ \ \ ;\nu}=0$ of the Einstein tensor 
when $\kappa\sim G/c^4$ is time-dependent.

\section{The Flatness Problem}\label{three}

In this section, we examine whether the flatness problem can be resolved by 
the bimetric model resulting from the brane world cosmology with the 
time-varying radion field.  The critical density $\varrho_c$, the mass 
density that gives rise to the flat universe ($k=0$) for a given 
$\dot{\bar{a}}_R/\bar{a}_R$, of the brane universe is given by
\begin{equation}
\varrho_c={3\over{8\pi\bar{G}_4}}\left({\dot{\bar{a}}_R\over\bar{a}_R}
\right)^2.
\label{critdens}
\end{equation}
We define the deviation of $\varrho$ from $\varrho_c$ as $\epsilon\equiv
\varrho/\varrho_c-1$.  Then, the $\epsilon<0$, $\epsilon=0$, and $\epsilon>0$ 
cases respectively correspond to the open ($k=-1$), flat ($k=0$) and closed 
($k=1$) universes.  In order for the flatness problem to be resolved, 
$\epsilon=0$ therefore has to be a stable attractor.  The time derivative of 
$\epsilon$ is
\begin{equation}
\dot{\epsilon}=2\epsilon\left({\dot{\bar{c}}_4\over\bar{c}_4}-
{\ddot{\bar{a}}_R\over\dot{\bar{a}}_R}\right),
\label{doteps}
\end{equation}
where $\dot{\bar{c}}_4/\bar{c}_4=e^{2m_0(R_0-R)}R_{\tau}R_{\tau\tau}/(c^2+
R^2_{\tau})$.  So, the rapid enough decrease of the speed of a graviton 
(\ref{effredeq3}) to the present value $c$ makes the terms in the bracket 
to be negative, causing $\epsilon=0$ to be an attractor.  This can be 
achieved by the radion potential with very steep region around the minimum, 
which allows rapid settling down of the radion to the minimum of the 
potential, causing sudden decrease of the speed of a graviton from a very 
large value to the present value.  

We now comment on the important difference of our case from Refs. 
\cite{cm1,cm2,blmv1,blmv2,cm3,cm4}.  It is claimed in Refs. \cite{blmv1,blmv2} 
that any bimetric implementation of cosmological models does not by itself 
solve the flatness problem.  The claim in Refs. \cite{blmv1,blmv2} is based 
on the fact that the speed of gravitons in their bimetric cosmological models 
is assumed to be constant (while the speed of photons varies with time) and 
therefore the first term in the bracket of Eq. (\ref{doteps}) vanishes.  So, 
in their bimetric models, the flatness problem can be solved only when 
$\ddot{\bar{a}}_R>0$, i.e., when the universe inflates by violating the strong 
energy condition.  Another important difference of the bimetric models 
considered in Refs. \cite{cm1,cm2,cm3,cm4} from our bimetric model is that, 
as we mentioned previously, in their case the speed of a photon varies with 
time, taking larger value than the speed of a graviton, or the speed of a 
graviton varies with time, taking smaller value than the speed of a photon, 
depending on the choice of the time coordinate, while the biscalar field 
varies with time.  So, in their case the speed of a graviton {\it increases} 
to the present day value while the biscalar field settles down to the minimum 
of the biscalar potential, causing the terms in the bracket of Eq. 
(\ref{doteps}) to be always positive unless the universe inflates rapidly  
enough.  This fact is in accordance with the claim in Refs. \cite{blmv1,blmv2} 
that the bimetric models of Refs. \cite{cm1,cm2,cm3,cm4} can solve the 
flatness problem only when the strong energy condition is violated
\footnote{On the other hand, it is claimed in Refs. \cite{cm3,cm4} that the 
flatness problem can be resolved because of the inflationary behavior of the 
bimetric models not due to an inflaton potential but due to the fact that the 
light cone of the matter metric was initially wider than that of the gravity 
metric and then contracted.  However, for our bimetric model, the light cone 
for the matter metric was initially narrower, so our bimetric model cannot 
solve the flatness problem in the manner described in Refs. \cite{cm3,cm4}.}.  
To sum up, the reason why our bimetric cosmological model resulting from the 
brane world scenario can solve the flatness problem, contrary to the negative 
claim in Refs. \cite{blmv1,blmv2}, is that in our case the speed of gravitons 
{\it varies} with time (unlike the models considered in \cite{blmv1,blmv2}) 
and becomes {\it larger} than the speed of photons (unlike the models 
considered in Refs. \cite{cm1,cm2,cm3,cm4}) while the biscalar or the radion 
varies with time.

\section{The Cosmological Constant Problem}\label{four}

In this section, we examine whether our bimetric model can solve the 
cosmological constant problem.  First of all, it is important to keep in 
mind that unlike the case of the conventional cosmology the mass density 
satisfying $\varrho=-\wp/\bar{c}^2_4$ is not directly related to the 
cosmological constant of the brane universe but rather to the brane tension.  
However, if we assume that the brane tensions initially took the fine-tuned 
values which give rise to zero effective four-dimensional cosmological 
constant and thereby the effective Friedmann equations of the forms 
(\ref{effecredeq3},\ref{fredeq2}) without cosmological constant term (of 
course, except for the radion potential term), then we can regard the 
nonzero mass density $\varrho_{\delta\sigma}$ satisfying $\varrho_{\delta
\sigma}=-\wp_{\delta\wp}/\bar{c}^2_4$ as being due to the correction $\delta
\sigma$ to the fine-tuned brane tensions which gives rise to nonzero 
effective four-dimensional cosmological constant in the brane universe.  

We now express the total mass density of the brane universe as the sum 
$\varrho=\varrho_m+\varrho_{\delta\sigma}$ of the mass density $\varrho_m$ of 
the ordinary matter fields on the brane and $\varrho_{\delta\sigma}=
\delta\sigma/\bar{c}^2_4$.  Then, the generalized conservation equation 
(\ref{genconseq}) is modified to
\begin{equation}
\dot{\varrho}_m+3\left(\varrho_m+{\wp_m\over\bar{c}^2_4}\right){\dot{\bar{a}}_R
\over\bar{a}_R}=-\dot{\varrho}_{\delta\sigma}-\varrho{\dot{\bar{G}}_4\over 
\bar{G}_4}+{{3k\bar{c}_4\dot{\bar{c}}_4}\over{4\pi\bar{G}_4\bar{a}^2_R}}.
\label{modgenconseq}
\end{equation}
The time derivative of the ratio $\epsilon_{\delta\sigma}=\varrho_{\delta
\sigma}/\varrho_m$ of $\varrho_{\delta\sigma}$ to $\varrho_m$ is given by
\begin{equation}
\dot{\epsilon}_{\delta\sigma}=\epsilon_{\delta\sigma}\left(
{\dot{\varrho}_{\delta\sigma}\over\varrho_{\delta\sigma}}-{\dot{\varrho}_m\over
\varrho_m}\right).
\label{ratdelsigm}
\end{equation}
Assuming that the brane matter satisfies the equation of state of the form 
$\wp_m=w\varrho_m\bar{c}^2_4$ with a constant $w$, we obtain
\begin{equation}
{\dot{\varrho}_{\delta\sigma}\over\varrho_{\delta\sigma}}=
-2{\dot{\bar{c}}_4\over\bar{c}_4},
\label{dotrat1}
\end{equation}
\begin{equation}
{\dot{\varrho}_m\over\varrho_m}=-3{\dot{\bar{a}}_R\over\bar{a}_R}(1+w)
-2{\dot{\bar{c}}_4\over\bar{c}_4}{\varrho_c\over\varrho_m}+2{\dot{\bar{c}}_4
\over\bar{c}_4}{{\varrho+\varrho_{\delta\sigma}}\over\varrho_m}
-{\varrho\over\varrho_m}{\dot{\bar{G}}_4\over\bar{G}_4}.
\label{dotrat2}
\end{equation}
Therefore, Eq. (\ref{ratdelsigm}) takes the form
\begin{equation}
\dot{\epsilon}_{\delta\sigma}=\epsilon_{\delta\sigma}\left[3{\dot{\bar{a}}_R
\over\bar{a}_R}(1+w)+2{\dot{\bar{c}}_4\over\bar{c}_4}{{1+
\epsilon_{\delta\sigma}}\over{1+\epsilon}}+\left({\dot{\bar{G}}_4\over
\bar{G}_4}-4{\dot{\bar{c}}_4\over\bar{c}_4}\right)(1+\epsilon_{\delta\sigma})
\right].
\label{ratdelsigm2}
\end{equation}
With the effective four-dimensional speed of a graviton and Newton's constant 
given in Eq. (\ref{effredeq3}), we have
\begin{equation}
{\dot{\bar{c}}_4\over\bar{c}_4}=e^{2m_0(R_0-R)}{{R_{\tau}R_{\tau\tau}}
\over{c^2+R^2_{\tau}}},
\label{dotratc}
\end{equation}
\begin{equation}
{\dot{\bar{G}}_4\over\bar{G}_4}=e^{2m_0(R_0-R)}{{R_{\tau}\left[3(\Omega^2_R-1)
R_{\tau\tau}+2m_0\Omega^2_R(c^2+R^2_{\tau})\right]}\over{(\Omega^2_R-1)
(c^2+R^2_{\tau})}}.
\label{dotratg}
\end{equation}
Making use of approximations $\Omega_R\approx 0$ and $e^{2m_0(R_0-R)}\approx 
1$ assumed in the RS models, we can put Eq. (\ref{ratdelsigm2}) into the form:
\begin{eqnarray}
\dot{\epsilon}_{\delta\sigma}&\approx&\epsilon_{\delta\sigma}\left[3
{\dot{\bar{a}}_R\over\bar{a}_R}(1+w)+(1+\epsilon_{\delta\sigma}){{1-\epsilon}
\over{1+\epsilon}}{{R_{\tau}R_{\tau\tau}}\over{c^2+R^2_{\tau}}}\right]
\cr
&\approx&\epsilon_{\delta\sigma}\left[3{\dot{\bar{a}}_R\over\bar{a}_R}(1+w)
+(1+\epsilon_{\delta\sigma}){{1-\epsilon}\over{1+\epsilon}}{\dot{\bar{c}}_4
\over\bar{c}_4}\right].
\label{ratdelsigm3}
\end{eqnarray}
So, as long as $\epsilon<1$, the rapid enough decrease of the speed of a 
graviton to the present value will cause the correction $\delta\sigma$ 
to the fine-tuned brane tensions to be driven to zero rapidly, thereby 
the brane tensions being pushed back to the fine-tuned values giving rise 
to zero cosmological constant in the brane universe.  
For the case when the speed of a graviton does not decease rapidly enough, 
$\delta\sigma$ is expected to approach a small constant value (while the 
radion is being stabilized) that gives rise to a small effective 
four-dimensional cosmological constant, solving the quasi cosmological 
constant problem in the manner proposed in Ref. \cite{bm}.  
Our bimetric model thereby can be used to bring the quantum corrections to 
the fine-tuned brane tensions after the SUSY breaking under control, just 
like the mechanism for self-tuning brane tension \cite{adk,kss}.  In this 
process, the correction $\delta\sigma$ to the fine-tuned to brane tensions 
is converted into ordinary matter.

Once again it is essential in this mechanism that the speed of a graviton 
varies with time, taking the value larger than the present value, while the 
radion varies with time.  The reason why the bimetric models in Refs. 
\cite{cm1,cm2,blmv1,blmv2,cm3,cm4} cannot solve the cosmological constant 
problem is that in their case the speed of a graviton either remains constant 
or takes smaller value than the present value while the biscalar varies with 
time.

\section{Other Cosmological Problems}\label{five}

The cosmic microwave background data indicates that photons emitted from 
the opposite sides of the sky appear to be in thermal equilibrium, although 
according to the SBB model those regions are out of causal contact at the 
time of last scattering, the so-called the horizon problem of the SBB model.  
The horizon problem is solved in inflationary models with sufficient amount 
of inflation, since the Hubble length measured in the comoving coordinates 
decreases during inflation.  The VSL models proposed in Refs. \cite{mof1,am} 
also solve the horizon problem, since the particle horizon scale at the last 
time of scattering $t_*$, given by
\begin{equation}
d_H(t_*)=\int^{t_*}_0{{c_p(t)dt}\over{a(t)}},
\label{horscat}
\end{equation}
can become larger than the coordinate distance to the last scattering, given 
by
\begin{equation}
d_H(t_*,t_0)=\int^{t_0}_{t_*}{{c_p(t)dt}\over{a(t)}},
\label{crdistprs}
\end{equation} 
where $t_0$ denotes the present epoch, if the speed of a photon $c_p$ were 
large enough during an initial period of cosmic evolution.  We have to keep in 
mind that the horizon scale is defined in terms of the speed of a photon 
$c_p$, instead of the speed of a graviton $c_g$, since we are considering the 
distance over which photons travel and transport energy.  Unfortunately, 
our bimetric model cannot solve the horizon problem through the mechanism 
proposed in Refs. \cite{mof1,am}, because the speed of a photon $c_p$ always 
remains constant with the choice of the time coordinate $\bar{\tau}$ or takes 
smaller value than the present day value $c$ with the choice of the comoving 
time coordinate for the gravity metric $g^{grav}_{\mu\nu}$, while the radion 
varies with time.  Therefore, we have to incorporate inflation into our 
bimetric model in order to solve the horizon problem.  Although an extra 
scalar field can be introduced as an inflaton, the radion may be used as an 
inflaton, if the radion potential has a region satisfying the slow-roll 
approximation.  

Modern particle theory models predict unwanted relics such as magnetic 
monopoles, domain walls, moduli fields, etc, during very early stage of 
cosmological evolution.  Since these relics get diluted more slowly than 
the (relativistic) ordinary matter as the universe expands, they become 
the dominant component of our present universe, which is in contraction 
with the observational data.   The problem of unwanted relics is solved 
in inflationary models if the temperature of the universe during the 
reheating was not high enough to produce the relics, because the relics, 
as well as matter, get diluted to the negligible level compared to the 
inflationary potential during the rapid expansion of the inflationary 
stage.  The VSL models can also solve the problem for the following reason.  
According to the Kibble mechanism, topological defect densities are 
inversely proportional to powers of the correlation length of the Higgs 
fields, which are generally bounded above by the Hubble distance $c_p/H$.  
So, if topological defects were created before the speed of a photon $c_p$ 
decreased to the present value, the upper bound on the densities of 
topological defects is weakened at the time of their production 
due to very large Hubble distance.  Since the mechanism for resolving the 
problem of the unwanted relics in the VSL models involves a larger value of 
speed of a photon during an early period, once again our bimetric model cannot 
by itself solve the problem.  The resolution of the problem therefore 
necessitates incorporation of other mechanism such as inflation.  

Finally, we comment on the explanation for the huge amount of entropy 
appears to be present in our universe.  The inflationary model explains 
the entropy problem by assuming that the adiabaticity condition is 
violated during the inflation:  While the universe supercools (due to the 
inflation) to some temperature $T_s$ and then reheats to $T_r$, the entropy 
density is increased by a factor of $(T_r/T_s)^3$.  In the case of the VSL 
models, the large production of entropy can be achieved through creation of 
particles while the speed of light, as well as the Newton's constant, 
changes to the present value, due to the nonconservation of the 
energy-momentum tensor (cf. see Refs. \cite{hm1,hm2}).  In order for particles 
to be created while the radion varies with time, the RHS of the generalized 
conservation equation (\ref{genconseq}) has to be positive.  Certainly, this 
can be achieve by the flat ($k=0$) and the open ($k=-1$) universes.  However, 
if the universe were closed ($k=1$) while the radion varies with time,  
particles would be take away, thereby entropy decreasing.  The production of 
sufficient amount of entropy may require the supplementation of our bimetric 
model by inflation.


\begin{thebibliography} {99}
\small
\parskip=0pt plus 2pt

\bibitem{mof1} J.W. Moffat, ``Superluminary universe: A Possible solution to 
the initial value problem in cosmology,'' Int. J. Mod. Phys. {\bf D2} (1993) 
351, gr-qc/9211020.

\bibitem{am} A. Albrecht and J. Magueijo, ``Time varying speed of light as 
a solution to cosmological puzzles,'' Phys. Rev. {\bf D 59} (1999) 043516, 
astro-ph/9811018.

\bibitem{gut} A.H. Guth, ``The inflationary universe: A possible solution to 
the horizon and flatness problems,'' Phys. Rev. {\bf D 23} (1981) 347.

\bibitem{lin} A.D. Linde, ``A new inflationary universe scenario: A possible 
solution of The horizon, flatness, homogeneity, isotropy and primordial 
monopole problems,'' Phys. Lett. {\bf B108} (1982) 389.

\bibitem{als} A. Albrecht and P.J. Steinhardt, ``Cosmology for grand unified 
theories with radiatively induced symmetry breaking,'' Phys. Rev. Lett. 
{\bf 48} (1982) 1220.

\bibitem{bar1} J.D. Barrow, ``Cosmologies with varying light speed,'' 
Phys. Rev. {\bf D 59} (1999) 043515, astro-ph/9811022.

\bibitem{bar2} J.D. Barrow and J. Magueijo, ``Varying-alpha theories and 
solutions to the cosmological problems,'' Phys. Lett. {\bf B443} (1998) 104, 
astro-ph/9811072.

\bibitem{bar3} J.D. Barrow and J. Magueijo, ``Solutions to the quasi-flatness 
and quasi-lambda Problems,'' Phys. Lett. {\bf B447} (1999) 246, 
astro-ph/9811073.

\bibitem{mof2} J.W. Moffat, ``Varying light velocity as a solution to the 
problems in cosmology,'' astro-ph/9811390.

\bibitem{bar4} J.D. Barrow and J. Magueijo, ``Solving the flatness and 
quasi-flatness problems in Brans-Dicke  cosmologies with a varying light 
speed,'' Class. Quant. Grav. {\bf 16} (1999) 1435, astro-ph/9901049.

\bibitem{wfc} J.K. Webb, V.V. Flambaum, C.W. Churchill, M.J. Drinkwater and 
J.D. Barrow, ``Evidence for time variation of the fine structure constant,'' 
Phys. Rev. Lett. {\bf 82} (1999) 884, astro-ph/9803165.

\bibitem{bek} J.D. Bekenstein, ``Fine structure constant: Is it really a 
constant?,'' Phys. Rev. {\bf D 25} (1982) 1527.

\bibitem{kst} M. Kaplinghat, R.J. Scherrer and M.S. Turner, ``Constraining 
variations in the fine-structure constant with the cosmic microwave 
background,'' Phys. Rev. {\bf D60} (1999) 023516, astro-ph/9810133.

\bibitem{bir} L. Bergstrom, S. Iguri and H. Rubinstein, ``Constraints on the 
variation of the fine structure constant from big  bang nucleosynthesis,'' 
Phys. Rev. {\bf D60} (1999) 045005, astro-ph/9902157.

\bibitem{bcw} R.A. Battye, R. Crittenden and J. Weller, ``Cosmic concordance 
and the fine structure constant,'' Phys. Rev. {\bf D63} (2001) 043505, 
astro-ph/0008265.

\bibitem{amrv} P.P. Avelino, C. J. Martins, G. Rocha and P. Viana, 
``Looking for a varying $\alpha$ in the Cosmic Microwave Background,'' 
Phys. Rev. {\bf D62} (2000) 123508, astro-ph/0008446.

\bibitem{mur} M.T. Murphy {\it et al.}, ``Possible evidence for a variable 
fine structure constant from QSO absorption lines: motivations, analysis and 
results,'' astro-ph/0012419.

\bibitem{web} J.K. Webb {\it et al.}, ``Further evidence for cosmological 
evolution of the fine structure constant,'' astro-ph/0012539.

\bibitem{ave2} P.P. Avelino {\it et al.}, ``Early-universe constraints on a 
time-varying fine structure constant,'' astro-ph/0102144.

\bibitem{kal1} G. Kalbermann and H. Halevi, ``Nearness through an extra 
dimension,'' gr-qc/9810083.

\bibitem{kir} E. Kiritsis, ``Supergravity, D-brane probes and thermal super 
Yang-Mills:  A comparison,'' JHEP{\bf 9910} (1999) 010, hep-th/9906206.

\bibitem{kal2} G. Kalbermann, ``Communication through an extra dimension,'' 
Int. J. Mod. Phys. {\bf A15} (2000) 3197, gr-qc/9910063.

\bibitem{chu} D.J. Chung and K. Freese, ``Can geodesics in extra dimensions 
solve the cosmological horizon  problem?,'' Phys. Rev. {\bf D 62} (2000) 
063513, hep-ph/9910235.

\bibitem{ale} S.H. Alexander, ``On the varying speed of light in a 
brane-induced FRW universe,'' JHEP{\bf 0011} (2000) 017, hep-th/9912037.

\bibitem{ish} H. Ishihara, ``Causality of the brane universe,'' Phys. Rev. 
Lett. {\bf 86} (2001) 381, gr-qc/0007070.

\bibitem{ckr} D.J. Chung, E.W. Kolb and A. Riotto, ``Extra dimensions present 
a new flatness problem,'' hep-ph/0008126.

\bibitem{csa} C. Csaki, J. Erlich and C. Grojean, ``Gravitational Lorentz 
violations and adjustment of the cosmological  constant in asymmetrically 
warped spacetimes,'' hep-th/0012143.

\bibitem{youm} D. Youm, ``Brane world cosmologies with varying speed of 
light,'' hep-th/0101228.

\bibitem{rs1} L. Randall and R. Sundrum, ``A large mass hierarchy from a 
small extra dimension,'' Phys. Rev. Lett. {\bf 83} (1999) 3370, 
hep-ph/9905221.

\bibitem{rs2} L. Randall and R. Sundrum, ``An alternative to 
compactification,'' Phys. Rev. Lett. {\bf 83} (1999) 4690, hep-th/9906064.

\bibitem{rs3} J. Lykken and L. Randall, ``The shape of gravity,'' JHEP 
{\bf 0006} (2000) 014, hep-th/9908076.

\bibitem{cm1} M.A. Clayton and J.W. Moffat, ``Dynamical mechanism for varying 
light velocity as a solution to cosmological problems,'' Phys. Lett. 
{\bf B460} (1999) 263, astro-ph/9812481.

\bibitem{cm2} M.A. Clayton and J.W. Moffat, ``Scalar-tensor gravity theory 
for dynamical light velocity,'' Phys. Lett. {\bf B477} (2000) 269, 
gr-qc/9910112.

\bibitem{dru} I.T. Drummond, ``Variable light-cone theory of gravity,'' 
gr-qc/9908058.

\bibitem{cgrt} C. Csaki, M. Graesser, L. Randall and J. Terning, 
``Cosmology of brane models with radion stabilization,'' Phys. Rev. {\bf D62} 
(2000) 045015, hep-ph/9911406.

\bibitem{cf2} J.M. Cline and H. Firouzjahi, ``5-dimensional warped 
cosmological solutions with radius stabilization  by a bulk scalar,'' Phys. 
Lett. {\bf B495} (2000) 271, hep-th/0008185.

\bibitem{bdl} P. Binetruy, C. Deffayet and D. Langlois, ``The radion in brane 
cosmology,'' hep-th/0101234.

\bibitem{gw1} W.D. Goldberger and M.B. Wise, ``Modulus stabilization with bulk 
fields,'' Phys. Rev. Lett. {\bf 83} (1999) 4922, hep-ph/9907447.

\bibitem{gw2} W.D. Goldberger and M.B. Wise, ``Phenomenology of a stabilized 
modulus,'' Phys. Lett. {\bf B475} (2000) 275, hep-ph/9911457.

\bibitem{cou} D.H. Coule, ``Varying $c$ cosmology has Planck epoch escape 
problem,'' Mod. Phys. Lett. {\bf A14} (1999) 2437, gr-qc/9811058.

\bibitem{blmv1} B.A. Bassett, S. Liberati, C. Molina-Paris and M. Visser, 
``Geometrodynamics of variable-speed-of-light cosmologies,'' Phys. Rev. 
{\bf D62} (2000) 103518, astro-ph/0001441.

\bibitem{blmv2} S. Liberati, B.A. Bassett, C. Molina-Paris and M. Visser, 
``$\chi$ variable-speed-of-light cosmologies,'' Nucl. Phys. Proc. Suppl. 
{\bf 88} (2000) 259, astro-ph/0001481.

\bibitem{cm3} M.A. Clayton and J.W. Moffat, ``Vector field mediated models 
of dynamical light velocity,'' gr-qc/0003070.

\bibitem{cm4} M.A. Clayton and J.W. Moffat, ``A scalar-tensor cosmological 
model with dynamical light velocity,'' gr-qc/0101126.

\bibitem{bm} J.D. Barrow and J. Magueijo, ``Solutions to the quasi-flatness 
and quasi-lambda problems,'' Phys. Lett. {\bf B447} (1999) 246, 
astro-ph/9811073.

\bibitem{adk} N. Arkani-Hamed, S. Dimopoulos, N. Kaloper and R. Sundrum, 
``A small cosmological constant from a large extra dimension,'' Phys. Lett. 
{\bf B480} (2000) 193, hep-th/0001197.

\bibitem{kss} S. Kachru, M. Schulz and E. Silverstein, ``Self-tuning flat 
domain walls in 5d gravity and string theory,'' Phys. Rev. {\bf D 62} (2000) 
045021, hep-th/0001206.

\bibitem{hm1} T. Harko and M.K. Mak, ``Particle creation in varying speed of 
light cosmological models,'' Class. Quant. Grav. {\bf 16} (1999) 2741.

\bibitem{hm2} M.K. Mak and T. Harko, ``Cosmological particle production in 
five-dimensional Kaluza-Klein  theory,'' Class. Quant. Grav. {\bf 16} (1999) 
4085.

\end{thebibliography}
\end{document}